\documentclass[final,epj]{svjour}
\usepackage{graphicx}
\usepackage[draft,implicit=false]{hyperref}
\usepackage{amsmath}

\begin{document}

\title{Weak measurement combined with quantum delayed-choice experiment and
implementation in optomechanical system}

\author{Gang Li,$^{1,*}$ Tao Wang,$^{2}$ Ming-Yong Ye,$^{3}$ and He-Shan Song$^{1,{\dag}}$}

\titlerunning{Weak measurement 
implementation in optomechanical system}

\authorrunning{Gang Li et al.}

\institute{$^{1}$School of Physics and Optoelectronic Technology, Dalian University of
Technology, Dalian 116024, China\\
$^{2}$College of Physics, Jilin University, Changchun 130012, China\\
$^{3}$School of Physics and Optoelectronics Technology, Fujian Normal University, Fuzhou 350007, China}
\mail{$^{*}$ligang0311@sina.cn, $^{\dag}$hssong@dlut.edu.cn}

\abstract{Weak measurement \cite{Aharonov88,Simon11} combined with quantum delayed-choice experiment that use quantum beam splitter instead of the beam splitter give rise to a surprising amplification effect, i.e., counterintuitive negative amplification effect. We show that this effect is caused by the wave and particle behaviours of the system to be and can't be explained by a semiclassical wave theory, due to the entanglement of the system and the ancilla in quantum beam splitter. The amplification mechanism about wave-particle duality in quantum mechanics lead us to a scheme for implementation of weak measurement in optomechanical system.}

\maketitle
\section{Introduction}

Postselection weak measurement was first proposed by Aharonov, Albert and
Vaidman \cite{Aharonov88}, where a pointer is weakly coupled to the system
to be measured. It can help us to understand some counter-intuitive quantum
paradoxes \cite{Aharonov05} and it is also used to measure and amplify small
physical quantities or effects which are not directly detected by
conventional techniques in experiment, such as direct measurement of wave
function \cite{Bamber11} and beam deflection \cite{Hosten08,Howell09}.
Recently, the basics and application of weak measurement have been reviewed
in \cite{Dressel14}.

Although postselection weak measurement have many applications, its
application in optomechanics was rarely reported. The optomechanical system
usually refers to a high finesse optical cavity with a tiny mirror attached
to micromechanical oscillator \cite{Girvin09,Marquardt13}, where the light
in the cavity can give a radiation force on the mirror. When there is only
one photon in the cavity the displacement of the mirror caused by the photon
is hard to be measured since it is much smaller than the spread of the
mirror wave packet. Recently, in the optomechanical system combined with
weak measurement, the weak measurement amplification scheme that the
displacement of the mirror caused by single photon can be amplified is
proposed \cite{Li14}, where the optomechanical system is embedded in the
March-Zehnder interferometer \cite{Scully97}. The amplification effect in
\cite{Li14} is obtained by retaining the Kerr phase in the Ref. \cite%
{Bouwmeester12}.

It is known to all that the Mach-Zehnder interferometer \cite{Scully97},
including the input beam splitter and the output beam splitter, can show
wave-particle duality. When a photon enters the input beam splitter, the
photon behaved as a wave behaviour if the output beam splitter is present or
behaved as a particle behaviour if the output beam splitter is absent. Based
on the Mach-Zehnder interferometer, the the delayed-choice experiment \cite%
{Wheeler78,Wheeler84} is proposed to demonstrate Bohr's complementarity \cite%
{Bohr84} and it show that wave-particle duality depend on the classical
detecting devices, i.e., whether the output beam splitter is inserted or
removed. However, recently Ionicioiu \emph{et al} propose a scheme for
quantum delayed-choice experiment \cite{Ionicioiu11} based on quantum beam
splitter in stead of the output beam splitter, where the quantum beam
splitter controlled by the ancilla \cite{Wheeler78,Wheeler84} can be put in
superposition of being present and absent. In this scheme, the
interferometer can be simultaneously closed and open so that both the wave
and particle behaviours of the photon can appear at the same time. Soon
after, this scheme is verified by a series of experiments \cite{Tang12,Peruzzo12,Kaiser12}.
\begin{figure}[bp]
\centering
\includegraphics[scale=0.48]{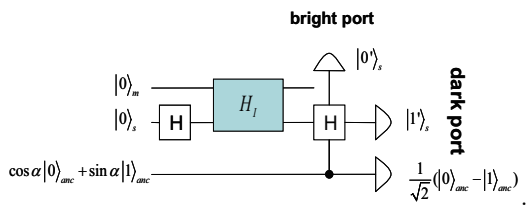}.
\caption{Weak measurement combined with quantum delayed-choice experiment. $
H_{I}$ is the weak interaction between the pointer and the system. Hadamard
gate is equivalent to the beam splitter. Controlled-Hadamard gate is
equivalent to quantum beam splitter and play a role of the postselection.
The ancilla is prepared in an arbitrary state $\cos \protect\alpha |0\rangle
_{anc}+\sin \protect\alpha |1\rangle _{anc}$}
\end{figure}

The nature of the quantum beam splitter is different from the beam splitter
(classical) and naturally we want to know based on quantum delayed-choice
experiment that use quantum beam splitter in stead of the beam splitter and
in combined with weak measurement whether there are some new features. In
the paper, we will first give a general discussion about weak measurements
\cite{Aharonov88,Simon11} combined with quantum delayed-choice experiment
that use the quantum beam splitter in stead of the beam splitter. The key
advantage of the ancilla in quantum beam splitter is to control the choice
of the measured system basis. When the quantum beam splitter possessing the
entanglement of the system and the ancilla play a role of the postselection,
we find that the wave and particle behaviours of the measured system give
rise to a surprising amplification effect and this amplification effect
can't be explained by Maxwell's equations \cite{Suter95}, due to the
entanglement of the system and the ancilla in quantum beam splitter. We
first show that the wave-particle duality can be used as the generation
mechanism for weak measurement amplification, that has not previously been
revealed. The scheme in this paper provide a new method for studying and
exploiting postselection and weak measurement \cite{Aharonov88,Simon11} in
three-qubit system. On the other hand, the perspective in this paper lead us
to a scheme for implementation of weak measurement \cite{Aharonov88,Simon11}
in optomechanical system. The amplification mechanism achieved here is
different from the one in \cite{Li14} which is caused by the Kerr phase.
Moreover, the displacement of the mirror's position in optomechanical system
can appear within a large evolution time zone, which is a counterintuitive
negative amplification effect.

The rest of this paper is organized in the following way. In Sec. 2, we give a general
discussion about weak measurements combined with quantum delayed-choice
experiment. In Sec. 3, we present a scheme for implementation of weak
measurement in optomechanical system, and In Sec. 4, we give the conclusion about
the work.

\section{Weak measurement amplification via quantum beam splitter}

To set the scene, we assume a general schematic diagram illustrated in Fig.
1. In the following the scheme involved a simple three-qubit system. The
pointer is a continuous system and initially prepared in the ground state $%
|0\rangle _{m}$. In the standard scenario of weak measurement, the
interaction Hamiltonian of the pointer and the system is expressed as
\begin{equation}
\hat{H}_{I}=\hbar \chi (t)\hat{\sigma}_{z}\otimes \hat{p},  \label{aa}
\end{equation}%
where $\hbar $ is Planck's constant, $\chi $ is a small coupling constant, $%
\sigma _{z}$ is a spin-like observable of the system to be measured and $%
\hat{p}$ is the momentum operator of the pointer.

Suppose that the state $|0\rangle _{s}$ is the initial state of the system,
where $|0\rangle _{s}$ is an eigenstate of $\hat{\sigma}_{z}$. Set $%
|+\rangle =\frac{1}{\sqrt{2}}(|0\rangle _{s}+|1\rangle _{s})$ and $\hat{%
\sigma}_{z}|+\rangle =|-\rangle $, where $|+\rangle $ and $|-\rangle $ is
eigenstates of $\hat{\sigma}_{x}$. The ancilla is an arbitrary state $\cos
\alpha |0\rangle _{anc}+\sin \alpha |1\rangle _{anc}$. Hadamard gate usually
plays a role of the beam splitters in quantum circuits \cite{Niesien00}. The
Hadamard transformation is
\begin{eqnarray}
|0^{^{\prime }}\rangle _{s} &=&\frac{1}{\sqrt{2}}(|0\rangle _{s}+|1\rangle
_{s}),  \notag \\
|1^{^{\prime }}\rangle _{s} &=&\frac{1}{\sqrt{2}}(|0\rangle _{s}-|1\rangle
_{s}),  \label{bb}
\end{eqnarray}
where $|0\rangle _{s}$ and $|1\rangle _{s}$ are two input modes entering the
beam splitter, $|0^{\prime }\rangle _{s}$ and $|1^{\prime }\rangle _{s}$ are
two output modes exiting the beam splitter. It is obvious that when Hadamard
gate is present, the measured system show wave behaviours. However, if
Hadamard gate is absent, the transformation between input and output modes
is
\begin{eqnarray}
|0^{\prime }\rangle _{s} &=&|0\rangle _{s},  \notag \\
|1^{\prime }\rangle _{s} &=&|1\rangle _{s},  \label{cc}
\end{eqnarray}%
showing particle behaviours. However in Ref. \cite{Ionicioiu11}, the
Controlled-Hadamard C(H) gate is equivalent to the quantum beam splitter,
where the quantum beam splitter controlled by ancilla \cite{Wheeler78} can
be put in superposition of being present and absent. In other word, the
state after quantum beam splitter in the interferometer is given by
\begin{equation}
|\psi _{QBS}\rangle =\cos \alpha |particle\rangle |0\rangle _{anc}+\sin
\alpha |wave\rangle |1\rangle _{anc},  \label{dd}
\end{equation}%
where $|particle\rangle =\frac{1}{\sqrt{2}}(|0^{\prime }\rangle
_{s}+|1^{\prime }\rangle _{s})$ and $|wave\rangle =|0^{\prime }\rangle _{s}$
describe the particle and wave behaviour states of the measured system,
respectively, and the ancilla $|0\rangle _{anc}$ and $|1\rangle _{anc}$
controls if the quantum beam splitter is absent and not.

Defining an annihilation operator $\hat{c}=\frac{1}{2\sigma }\hat{q}+i\frac{%
\sigma }{\hbar }\hat{p}$, where $\hat{q}$ is position operator (the
canonical variable conjugates to $\hat{p}$ and $[q,p]=i\hbar $). According
to the result of Ref. \cite{Simon11}, the Hamiltonian above can be rewritten
as
\begin{equation}
\hat{H}_{I}=-i\frac{\hbar ^{2}\chi (t)}{2\sigma }\hat{\sigma}_{z}(\hat{c}-%
\hat{c}^{\dagger }),  \label{ee}
\end{equation}%
where $\sigma $ is the zero-point fluctuation and $\sqrt{2}\sigma $ is the
width of the Gaussian distribution. Then the time evolution of the system
and the pointer is given by
\begin{eqnarray}
e^{-\frac{i}{\hbar }\int \hat{H}dt}|+\rangle |0\rangle &=&\exp [-\eta \hat{%
\sigma}_{z}(\hat{c}-\hat{c}^{\dagger })]|+\rangle _{s}|0\rangle _{m}  \notag
\\
&=&\frac{1}{\sqrt{2}}(|0\rangle _{s}D(\eta )|0\rangle _{m}  \notag \\
&+&|1\rangle _{s}D(-\eta )|0\rangle _{m}),  \label{ff}
\end{eqnarray}%
where $D(\eta )=\exp [(\eta \hat{c}^{\dagger }-\eta ^{\ast }\hat{c})]$ with $%
\eta =\frac{\hbar \chi }{2\sigma }$ is a displacement operator and $\eta \ll
1$. After the quantum beam splitter and using the transformation of Eq. (\ref%
{bb}) and Eq. (\ref{cc}), then the state of the total system becomes
\begin{eqnarray}
&&\frac{\cos \alpha }{\sqrt{2}}(|0\rangle _{s}D(\eta )|0\rangle
_{m}+|1\rangle _{s}D(-\eta )|0\rangle _{m})|0\rangle _{anc}  \notag \\
&+&\frac{\sin \alpha }{2}[(|+\rangle (D(\eta )|0\rangle _{m}+D(-\eta
)|0\rangle _{m})  \notag \\
&+&|-\rangle (D(\eta )|0\rangle _{m}-D(-\eta )|0\rangle _{m})]|1\rangle
_{anc}.  \label{gg}
\end{eqnarray}%
When the state $|1^{\prime }\rangle $ is detected, in the language of weak
measurement we simultaneously measure the system in the orthogonal
postselection state of wave behaviours $|-\rangle $ \cite{Bouwmeester12,Li14}
and in the state of the particle behaviour $|1\rangle _{s}$ before the
ancilla, i.e., before choosing if quantum beam splitter is present or
absent, then the state of the pointer and the ancilla becomes
\begin{eqnarray}
|\psi \rangle _{m} &=&\frac{\cos \alpha }{\sqrt{2}}D(-\eta )|0\rangle
_{m})|0\rangle _{anc}  \notag \\
&+&\frac{\sin \alpha }{2}(D(\eta )|0\rangle _{m}-D(-\eta )|0\rangle
_{m})]|1\rangle _{anc}.  \label{hh}
\end{eqnarray}%
Results indicate that even though the system is measured, we can vary the
system from wave to particle behaviour through adjusting the angle $\alpha $
in ancilla. It is interesting to consider the fact that when the ancilla is
measured in a superposition state, such as $|\varphi \rangle =\frac{1}{\sqrt{%
2}}(|0\rangle _{anc}-|1\rangle _{anc})$, then the final state of the mirror
is a superposition state,
\small
\begin{figure}[t]
\includegraphics[scale=0.4]{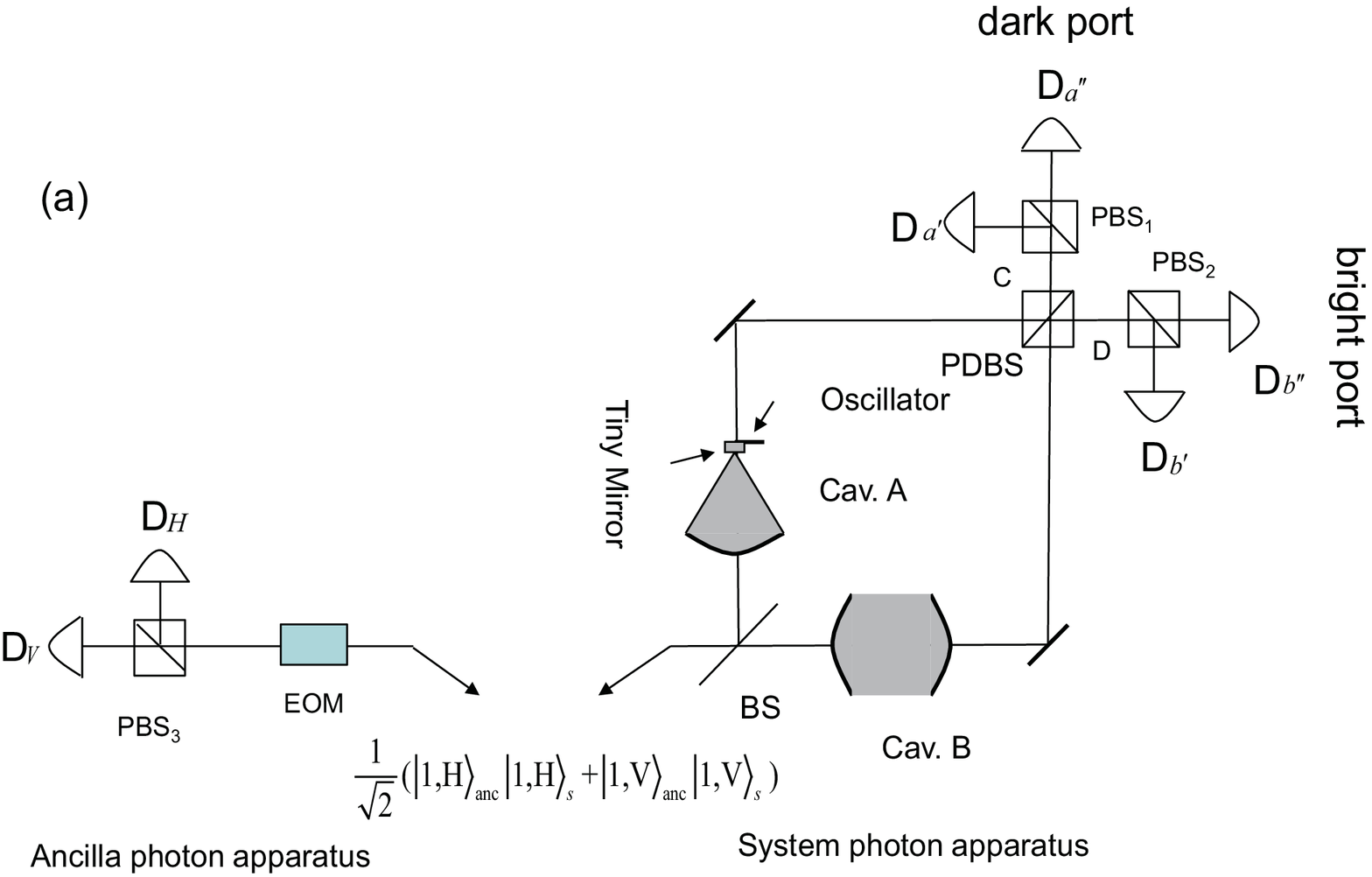}~\newline
\includegraphics[scale=0.4]{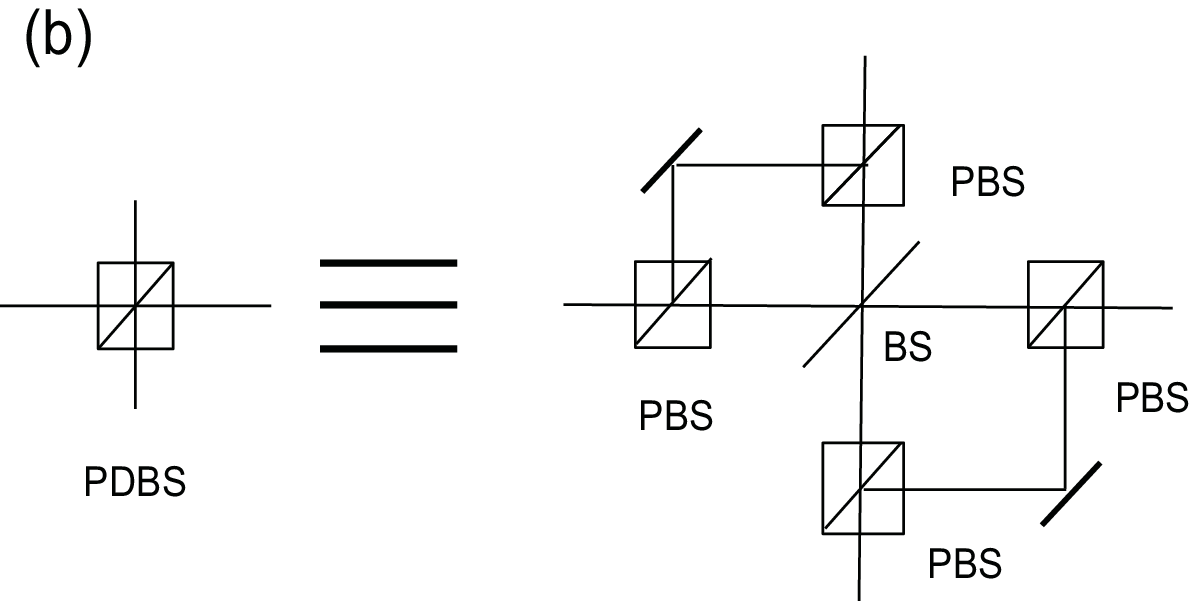}.
\caption{(a) Quantum delayed-choice experiment in optomechanical system. The
one of the entangled photons enters the first beam splitter of March-Zehnder
interferometer with an optomechanical cavity A and a conventional cavity B
on the right side. The quantum beam splitter consist of the
polarization-dependent beam splitter (PDBS) and the erasure device, i.e.,
the polarizing beam splitter (PBS$_{1}$ and PBS$_{2}$). The ancilla photon
of the entangled state is sent to ancilla photon apparatus on the left side
where EOM denote electro-optic phase modulator and PBS$_3$ is the polarizing
beam splitter oriented at the \{$H$,$V$\} basis (b) Polarization-dependent
beam splitter (PDBS) yield the $100/0$ reflection/transimission ratio for $%
|H\rangle$ and $50/50$ reflection/transimission ratio for $|V\rangle$. In
PDBS, the four polarizing beam-splitter (PBS) orient at the \{$H$,$V$\}
basis and the BS is an ordinary $50/50$ BS.}
\end{figure}

\begin{eqnarray}
|\psi \rangle _{m} &=&\frac{\cos \alpha }{\sqrt{2}}D(-\eta )|0\rangle _{m}
\notag \\
&-&\frac{\sin \alpha }{2}(D(\eta )|0\rangle _{m}-D(-\eta )|0\rangle _{m})].
\label{aaa}
\end{eqnarray}%
We note that $\cos \alpha $ is caused by the particle behaviour of the
system, while $\sin \alpha $ is due to wave behaviour of the system. From
the derivation of Eq. (\ref{aaa}), it will be shown that the key advantage
of the ancilla measurement is to control the choice of the measured system
basis. This indicate that the Controlled-Hadamard gate or quantum beam
splitter play a role of the postselection. Moreover, the state of Eq. (\ref%
{aaa}) can also be obtained when we simultaneously measure the system and
the ancilla in the entangled states, i.e., the postselection state
\begin{equation}
|\psi _{f}\rangle =|1\rangle _{s}|0\rangle _{anc}-|-\rangle |1\rangle _{anc}.
\label{bbb}
\end{equation}
Therefore, the superposition of the pointer in Eq. (\ref{aaa}) caused by the
particle and wave behaviours of the system is no classical analog.

Based on the Eq. (\ref{aaa}), we can then perform a small quantity expansion
about $\eta $ till the second order, where $\eta \ll 1$, then
\begin{equation}
|\psi \rangle _{m}\approx \frac{\cos \alpha }{\sqrt{2}}(|0\rangle _{m}+\eta
|1\rangle _{m})-\eta \sin \alpha |1\rangle _{m}.  \label{ccc}
\end{equation}%
For $\alpha =\pi /2$, the beam splitter is present, i.e., only
wave behaviour occur, and we find the state $|\psi \rangle _{m}$
proportional to $|1\rangle _{m}$, which correspond to the case that the
post-selected state of the system is orthogonal to the initial state of the
system \cite{Aharonov88,Simon11}, the displacement of pointer's position is
zero.

For $\pi /2-\alpha \ll 1$, the particle and wave behaviour simultaneously
occur, here we use the approximation $\cos \alpha \approx \pi /2-\alpha $
and $\sin \alpha \approx 1$ to get the superposition state
\begin{equation}
|\psi \rangle _{m}\approx (\pi /2-\alpha )/\sqrt{2}|0\rangle _{m}-\eta
|1\rangle _{m},  \label{ddd}
\end{equation}%
where the coefficient $(\pi /2-\alpha )/\sqrt{2}$ is due to $\cos \alpha $
generated by the particle behaviour of the system, while the coefficient $%
\eta $ is due to $\eta \sin \alpha $ caused by the wave behaviour of the
system.

The average displacement of the pointer position $\hat{q}$ is
\begin{equation}
\langle \hat{q}\rangle =\frac{\langle \psi |_{m}\hat{q}|\psi \rangle _{m}}{%
\langle \psi |\psi \rangle _{m}}-\langle 0|\hat{q}|0\rangle .  \label{eee}
\end{equation}%
For Eq. (\ref{ddd}), it can be seen that $\langle \hat{q}\rangle $ is
none-zero (amplification) since the superposition between the states $%
|0\rangle _{m}$ and $|1\rangle _{m}$. Such a result stems from the wave and
particle behaviours of the system controlled by quantum beam splitter that
play a role of the postselection. It will be shown that wave-particle
duality in quantum mechanics can be used as generation mechanism of this
amplification effect, that has never before been revealed in weak
measurement. Moreover, the weak measurement provided here relies on the
entanglement of the system and the ancilla in quantum beam splitter, mean
that the results can't be explained by Maxwell's equations alone \cite%
{Suter95}. This scheme provide a new method for studying and exploiting
postselection and weak measurement in three-qubit system.

\section{ Implementation in optomechanical system}

In the following we will consider an weak measurement model in
optomechanical system combined with quantum delayed-choice experiment.

\subsection{ The optomechanical model}

Based on a Mach-Zehnder interferometer, a schematic diagram of quantum
delayed-choice experiment to be considered is shown in Fig. 2. The
optomechanical cavity A is embedded in one arm of the March-Zehnder
interferometer and a stationary Fabry-P \'{e}rot cavity B is placed in another
arm. And the first beam splitter is symmetric, but the second one is quantum
beam splitter \cite{Kaiser12} which is made up of two components. The first
is a polarization-dependent beam splitter (PDBS) which shows $100\%$
reflection for horizontal polarized photons and $50/50$
relection/transmission for vertical polarized photons. The second is erasure
device consisting of polarizing beam splitters (PBS$_{1}$ and PBS$_{2}$). In
Fig. 2 on the left side, the electro-optic phase modulator is to rotate the
polarization state of the the ancilla's photon by an angle $\alpha$.

The Hamiltonian of optomechanical system in the interferometer is expressed
as followed:
\begin{equation}
H=\hbar \omega _{0}(a^{\dagger }a+b^{\dagger }b)+\hbar \omega _{m}c^{\dagger
}c-\hbar ga^{\dagger }a(c^{\dagger }+c),  \label{fff}
\end{equation}%
where $\hbar $ is Planck's constant, $\omega _{0}$ is frequency of the
optical cavity A, B and the corresponding annihilation operators are $\hat{a}
$ and $\hat{b}$, $\omega _{m}$ is frequency of mechanical system and the
corresponding annihilation operator is $\hat{c}$, and the optomechanical
coupling strength $g=\frac{\omega _{0}}{L}\sigma $, where $L$ is the length
of the cavity A or B, $\sigma =(\hbar /2m\omega _{m})^{1/2}$ which is the
zero-point fluctuation and $m$ is the mass of mechanical system. Here it is
a weak measurement model where the mirror is used as the pointer to measure
the number of photon in cavity A. Noted that because $a^{\dag }a$ of the Eq.
(\ref{fff}) corresponds to $\hat{\sigma}_{z}$ of the Hamiltonian in the
standard scenario of weak measurement and $c+c^{\dag }$ corresponds to $\hat{%
p}$.

In the optomechanical cavity A (see Fig. 2), if the initial state of the
mirror is prepared at the ground state $|0\rangle $ and when one photon
interacts with the mirror, the mirror will become a coherent displacement
state \cite{Mancini97,Bose97}, $|\xi (t)\rangle =e^{i\phi (t)}|\varphi
(t)\rangle _{m}$, where $e^{i\phi (t)}$ is the Kerr phase of one photon with
$\phi (t)=k^{2}(\omega _{m}t-\sin \omega _{m}t)$, $\varphi
(t)=k(1-e^{-i\omega _{m}t})$ and $k=g/\omega _{m}$. According to the
expression of the position displacement $\langle \xi (t)|\hat{q}|\xi
(t)\rangle -\langle 0|\hat{q}|0\rangle $, then the displacement of the
mirror is not more than $4k\sigma $ for any time $t$ and can not be bigger
than the zero-point fluctuation of the mirror since $k=g/\omega _{m}$ can
not be bigger than $0.25$ in weak coupling condition \cite{Marshall03}.
Therefore, the displacement of the mirror caused by one photon can not be
detected. However, if considering optomechanical cavity with weak
measurement combined with quantum delayed-choice experiment, we can amplify
the mirror's displacement.
\begin{figure}[t]
\includegraphics[scale=0.46]{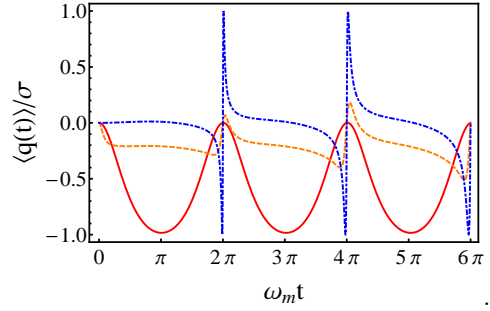}.
\caption{$\langle q(t)\rangle /\protect\sigma $ as a function of $\protect%
\omega _{m}t$ with $k=0.005$ for different $\protect\alpha $, i.e, $\protect%
\alpha =0.996\times \protect\pi /2$ (red-solid), $0.9995\times \protect\pi %
/2 $ (orange-dashed) and $\protect\pi /2$ (blue-dotted-dashed).}
\end{figure}

\subsection{Amplification with postselected weak measurement in optomechanics}

Suppose the one of the photons from the maximally polarization-entangled
Bell state $|\Phi ^{+}\rangle =\frac{1}{\sqrt{2}}(|1,H\rangle
_{anc}|1,H\rangle _{s}+|1,V\rangle _{anc}|1,V\rangle _{s})$ is input into
the interferometer, where $H$ ($V$) represent horizontal (vertical)
polarized photonic modes, then the state of the photons after the first beam
splitter becomes
\begin{eqnarray}
|\psi _{i}\rangle &=&\frac{1}{2}[|1,H\rangle _{anc}(|1,H\rangle
_{A}+|1,H\rangle _{B})  \notag \\
&&+|1,V\rangle _{anc}(|1,V\rangle _{A}+|1,V\rangle _{B})].  \label{ggg}
\end{eqnarray}%
After interacting weakly with the mirror prepared at ground state $|0\rangle
_{m}$ that can be achieved using sideband-resolved cooling technique \cite%
{Kippenberg07,Girvin07}, the state of the photons and the mirror is
\begin{eqnarray}
|\psi _{i}\rangle &=&\frac{1}{2}[|1,H\rangle _{anc}(|1,H\rangle _{A}|\xi
(t)\rangle _{m}+|1,H\rangle _{B}  \notag \\
&&|0\rangle _{m})+|1,V\rangle _{anc}(|1,V\rangle _{A}|\xi (t)\rangle _{m}
\notag \\
&&+|1,V\rangle _{B}|0\rangle _{m})].  \label{hhh}
\end{eqnarray}%
After the photon passes through the polarization-dependent beam splitter
(PDBS), the state of the overall system becomes
\begin{eqnarray}
|\psi _{PDBS}\rangle &=&\frac{1}{2}|1,H\rangle _{anc}(|1,H\rangle _{C}|\xi
(t)\rangle _{m}+|1,H\rangle _{D}  \notag \\
&&|0\rangle _{m})+\frac{1}{2\sqrt{2}}|1,V\rangle _{anc}[|1,V\rangle
_{C}(-|\xi (t)\rangle _{m}  \notag \\
&&+|0\rangle _{m})+|1,V\rangle _{D}(|\xi (t)\rangle _{m}+|0\rangle _{m})].
\label{iii}
\end{eqnarray}

In order to erasure of all polarization information about PDBS output, we
make the state of Eq. (\ref{iii}) go through the polarizing beam splitter
(PBS$_{1}$ and PBS$_{2}$) oriented at $45^{\circ }$ to \{$H$, $V$\} basis.
Then Eq. (\ref{iii}) becomes
\begin{equation}
|\psi \rangle =\frac{1}{\sqrt{2}}(|1,H\rangle _{anc}|particle\rangle
_{sm}+|1,V\rangle _{anc}|wave\rangle _{sm})  \label{jjj}
\end{equation}%
with
\begin{eqnarray}
|particle\rangle _{sm} &=&\frac{1}{2}(|1\rangle _{a^{^{\prime }}}+|1\rangle
_{a^{^{\prime \prime }}})|\xi (t)\rangle _{m}+(|1\rangle _{b^{^{\prime }}}
\notag \\
&&+|1\rangle _{b^{^{\prime \prime }}})|0\rangle _{m}  \label{kkk}
\end{eqnarray}%
and
\begin{eqnarray}
|wave\rangle _{sm} &=&\frac{1}{2\sqrt{2}}(-|1\rangle _{a^{^{\prime
}}}+|1\rangle _{a^{^{\prime \prime }}})(-|\xi (t)\rangle _{m}+|0\rangle _{m})
\notag \\
&&+(-|1\rangle _{b^{^{\prime }}}+|1\rangle _{b^{^{\prime \prime }}})(|\xi
(t)\rangle _{m}+|0\rangle _{m}).  \label{lll}
\end{eqnarray}

Using the electro-optic phase modulator (EOM) rotate the polarization state
of ancilla by an angle $\alpha $, we can obtain
\begin{eqnarray}
|\psi _{QBS}\rangle &=&\frac{1}{\sqrt{2}}[|1,H\rangle _{anc}(\cos \alpha
|particle\rangle _{sm}-\sin \alpha  \notag \\
&&|wave\rangle _{sm})+|1,V\rangle _{anc}(\cos \alpha |wave\rangle _{sm}
\notag \\
&&+\sin \alpha |particle\rangle _{sm})]  \label{mmm}
\end{eqnarray}

Based on the polarizing beam splitter (PBS$_{3}$) and detecting the
ancilla's photon in D$_{H}$, then we find that
\begin{equation}
|\psi _{QBS}\rangle =\frac{1}{\sqrt{2}}(\cos \alpha |particle\rangle
_{sm}-\sin \alpha |wave\rangle _{sm})  \label{nnn}
\end{equation}
Here we only consider detector D$_{a^{^{\prime }}}$. When a photon is
detected at the dark port D$_{a^{^{\prime }}}$, in the language of weak
measurement we simultaneously measure the single-photon in the orthogonal
postselection state of wave behaviours $\frac{1}{\sqrt{2}}(|1\rangle
_{A}-|1\rangle _{B})$ \cite{Bouwmeester12,Li14} and in the state of the
particle behaviour $|1\rangle _{A}$. Then the state of the mirror after
postselection is
\begin{equation}
|\psi _{m}\rangle =\frac{\cos \alpha }{2\sqrt{2}}|\xi (t)\rangle _{m}-\frac{%
\sin \alpha }{4}(|\xi (t)\rangle _{m}-|0\rangle _{m}).  \label{bbbb}
\end{equation}%
We note that $\cos \alpha $ is due to the particle behaviour of the photon,
while $\sin \alpha $ is due to wave behaviour of the photon.

The average displacement of the mirror's position operator $\hat{q}=(\hat{c}+%
\hat{c}^{\dag })\sigma $ is given by
\begin{equation}
\langle q\rangle =\frac{Tr(|\psi _{m}\rangle \langle \psi _{m}|q)}{Tr(|\psi
_{m}|\rangle \langle \psi _{m}|)}-Tr(|0\rangle _{m}\langle 0|_{m}q).
\label{dddd}
\end{equation}%
Substituting Eq. (\ref{bbbb}) into Eq. (\ref{dddd}), then
\begin{eqnarray}
\langle q(t)\rangle &=&\sigma \lbrack (2-\sin ^{2}\alpha -\sqrt{2}\sin
2\alpha )(\varphi (t)+\varphi ^{\ast }(t))  \notag \\
&+&(\sin 2\alpha /\sqrt{2}-\sin ^{2}\alpha )e^{-\frac{|\varphi (t)|^{2}}{2}%
}(e^{i\phi (t)}\varphi (t)  \notag \\
&+&e^{i\phi ^{\ast }(t)}\varphi ^{\ast }(t))]/[2-\sqrt{2}\sin 2\alpha +(\sin
2\alpha /\sqrt{2}  \notag \\
&+&\sin ^{2}\alpha )(e^{i\phi (t)}+e^{i\phi ^{\ast }(t)})]  \label{eeee}
\end{eqnarray}

To see this clearly, in Fig. 3, we plot the average displacement $\langle
q(t)\rangle /\sigma $ of the mirror versus the time $\omega _{m}t$ with $%
k=0.005$ for different angles $\alpha $ [$\alpha =0.996\times \pi /2$
(red-solid), $0.9995\times \pi /2$ (orange-dashed) and $\pi /2$
(blue-dotted-dashed)]. It can be seen clearly from Fig. 2 that when $\alpha
=0.996\times \pi /2$ the amplification (red-solid) is a time function of
period $2\pi $ and at time $\omega _{m}t=(2n+1)\pi $ $(n=0,1,2,\cdots )$ the
amplifications can reach strong coupling limits (the level of the ground
state fluctuation) $\langle q\rangle =-\sigma $ \cite{Marshall03}. However,
when $\alpha \rightarrow \pi /2$, such as $\alpha =0.9995\times \pi /2$, the
amplification around time $\omega _{m}t=2n\pi $ $(n=1,2,\cdots )$ appears,
but the rest of the amplifications (orange-dashed) is decreasing. Until $%
\alpha =\pi /2$ which indicates that the quantum beam splitter corresponds
to a beam splitter, the amplification (blue-dotted-dashed) around time $%
\omega _{m}t=2n\pi $ $(n=1,2,\cdots )$ is very prominent and reaches the
level of the ground state fluctuation $\langle q\rangle =\pm \sigma $. The
result is caused by the kerr phase which can't be explained by the standard
weak measurement \cite{Aharonov88,Simon11} and has been discussed in detail
in Ref. \cite{Li14}. Note that in the optomechanical cavity A (see Fig. 2)
the maximal displacement of the mirror caused by one photon is $4k\sigma $
and the displacement $-\sigma $ can be obtained using quantum beam splitter
in weak measurement, hence the amplification factor can be $Q=1/4k$ which is
$-50$ when $k=0.005$. Moreover, the negative amplification is a
counterintuitive result since intuitively there cannot exist negative
displacement relative to the direction of the photon propagation.
\begin{figure}[t]
\includegraphics[scale=0.46]{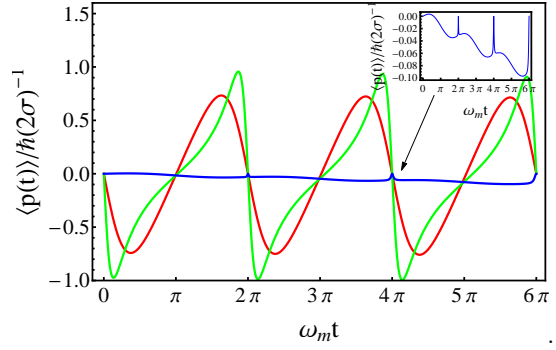}.
\caption{$\langle p(t)\rangle /\hbar (2\protect\sigma )^{-1}$ as a function
of $\protect\omega _{m}t$ with $k=0.005$ for different $\protect\alpha $,
i.e., $\protect\alpha =0.996\times \protect\pi /2$ (red), $0.999\times
\protect\pi /2$ (green) and $\protect\pi /2$ (blue).}
\end{figure}

\subsection{ Small quantity expansion about time for amplification}

To explain the amplification phenomenon shown in Fig. 3, we only give an
analysis about the special case that the maximal amplification appearing at
the time $T=(2n+1)\pi $ $(n=0,1,2,\cdots )$. In the same way as the Eq. (\ref%
{ddd}), for Eq. (\ref{bbbb}) we can perform a small quantity expansion about
time $T$ till the second order. Suppose that $|\omega _{m}t-T|\ll 1$, and $%
k\ll 1$, then
\begin{equation}
|\psi _{m}\rangle \approx \frac{\cos \alpha }{\sqrt{2}}(|0\rangle
_{m}+2k|1\rangle _{m})-k\sin \alpha |1\rangle _{m}.  \label{ffff}
\end{equation}%
When $\pi /2-\alpha \ll 1$, here we use the approximation $\cos \alpha
\approx \pi /2-\alpha $ and $\sin \alpha \approx 1$ to get the superposition
state
\begin{equation}
|\psi _{m}\rangle \approx (\pi /2-\alpha )/\sqrt{2}|0\rangle _{m}-k|1\rangle
_{m}.  \label{gggg}
\end{equation}%
Noted that the coefficient $(\pi /2-\alpha )/\sqrt{2}$ is due to $\cos
\alpha $ by generated the particle behaviour of the photon, while the
coefficient $k$ is due to $k\sin \alpha $ caused by the wave behaviour of
the photon. Substituting Eq. (\ref{gggg}) into Eq. (\ref{dddd}), the average
value of the displacement operator $\hat{q}$ is given by
\begin{eqnarray}
\langle q(t)\rangle _{|\omega _{m}t-T|\ll 1} &=&-\sigma \sqrt{2}k(\pi
/2-\alpha )/[k^{2}  \notag \\
&+&(\pi /2-\alpha )/2],  \label{hhhh}
\end{eqnarray}%
which can get its minimal value $-\sigma $ when $k=(\pi /2-\alpha )/\sqrt{2}$%
. Hence the state of the mirror achieving the maximal negative amplification
is $\frac{1}{\sqrt{2}}(|0\rangle _{m}-|1\rangle _{m})$. The results show
that the key to the amplification is the superposition of the vacuum state
and one phonon state of the mirror, which is due to the superposition of
wave and particle behaviours of the photon. Such result has not previously
been revealed. For Eq. (\ref{gggg}), the amplification mechanism is accord
with standard weak measurement \cite{Aharonov88,Simon11} and it is caused by
proper near-orthogonal postselection but different from the amplification
mechanism for Eq. (16) in Ref. \cite{Li14} which is caused by the Kerr phase
in this case of orthogonal postselection.

Next we would like to discuss the amplification of momentum variable $p$ of
the mirror. Substituting Eq. (\ref{bbbb}) into Eq. (\ref{dddd}) which uses $%
p $ $=-i(c-c^{\dagger })\frac{\hbar }{2\sigma }$ instead of $q$, then we
have
\begin{eqnarray}
\langle p(t)\rangle &=&-i\frac{\hbar }{2\sigma }[(2-\sin ^{2}\alpha -\sqrt{2}%
\sin 2\alpha )(\varphi (t)-\varphi ^{\ast }(t))  \notag \\
&+&(\sin 2\alpha /\sqrt{2}-\sin ^{2}\alpha )e^{-|\varphi (t)|^{2}}(e^{i\phi
(t)}\varphi (t)  \notag \\
&-&e^{i\phi ^{\ast }(t)}\varphi ^{\ast }(t))]/[2-\sqrt{2}\sin 2\alpha +(\sin
2\alpha /\sqrt{2}  \notag \\
&-&\sin ^{2}\alpha )(e^{i\phi (t)}+e^{i\phi ^{\ast }(t)})]  \label{iiii}
\end{eqnarray}%
The average displacement $\langle p(t)\rangle /\frac{\hbar }{2\sigma }$ of
the mirror is shown in Fig. 4 as a function of $\omega _{m}t$ with $k=0.005$
for different angle $\alpha $ [$\alpha =0.996\times \pi /2$ (red-solid), $%
0.999\times \pi /2$ (green-solid) and $\pi /2$ (blue-solid)]. We can see
clearly from Fig. 4 that the amplification effect (green-solid) around the
vibration periods of the mirror $\omega _{m}t=2n\pi $ $(n=0,1,2,\cdots )$ is
very prominent and reaches strong coupling limits (the level of the ground
state fluctuation) $\langle p\rangle =\pm \frac{\hbar }{2\sigma }$ \cite%
{Marshall03} if the angle $\alpha $ is very close to $\pi /2$. However, when
$\alpha =\pi /2$ the amplification of the mirror's momentum $p$ (blue-solid)
is suppressed around time $\omega _{m}t=2n\pi $ $(n=0,1,2,\cdots )$ (see the
inset in Fig. 4).

In order to understand the amplification phenomenon around time $\omega
_{m}t=2n\pi $ $(n=0,1,2,\cdots )$, for Eq. (\ref{bbbb}) we can also make a
small quantity expansion about time $T=2n\pi $ $(n=0,1,2,\cdots )$ till the
second order. Suppose that $\pi /2-\alpha \ll 1$, $|\omega _{m}t-T|\ll 1$
and $k\ll 1$, then
\begin{equation}
|\psi _{m}\rangle \approx (\pi /2-\alpha )/\sqrt{2}|0\rangle _{m}-ik(\omega
_{m}t-T)|1\rangle _{m}.  \label{jjjj}
\end{equation}%
It is obvious from the Eq. (\ref{jjjj}) that the key to the amplification is
also the superposition of the vacuum state and one phonon state of the
mirror, which is due to the superposition of wave and particle behaviours of
the photon. It is not hard to see that its maximal value is $\frac{\hbar }{%
2\sigma }$ when $(\pi /2-\alpha )/\sqrt{2}=-k(\omega _{m}t-T)$, and its
minimal one is $-\frac{\hbar }{2\sigma }$ when $(\pi /2-\alpha )/\sqrt{2}%
=k(\omega _{m}t-T)$. Therefore the mirror state achieving the maximal
positive amplification is $\frac{1}{\sqrt{2}}(|0\rangle _{m}+|1\rangle _{m})$
and the state achieving the maximal negative amplification is $\frac{1}{%
\sqrt{2}}(|0\rangle _{m}-|1\rangle _{m})$.
\begin{figure}[t]
\includegraphics[scale=0.43]{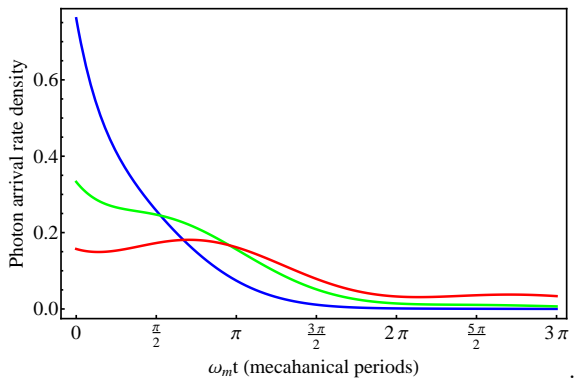}.
\caption{For $k=0.005$ and $\protect\alpha=0.996\times\protect\pi/2$, photon
arrival probability density vs arrival time for a successful postselection
with $\protect\omega_{m}=\protect\kappa$ (blue), $\protect\omega_{m}=2%
\protect\kappa$ (green) and $\protect\omega_{m}=4\protect\kappa$ (red).}
\end{figure}

\subsection{Discussion}

We shall show that how the photon arrival rates vary with time after the
single photon is detected at dart port. The probability density of a photon
being released from optomechanical cavity after time $\omega _{m}t$ is
\begin{equation}
\kappa \exp (-\kappa t),  \label{kk}
\end{equation}
where $\kappa $ is the decay rate of the cavity. The probability of a
successful post-selection being released after $\omega _{m}t$ is
\begin{eqnarray}
&&\frac{1}{4}[2-\sqrt{2}\sin 2\alpha +(\sin 2\alpha /\sqrt{2}-\sin
^{2}\alpha )(e^{i\phi (t)}  \notag \\
&&+e^{i\phi ^{\ast }(t)})].  \label{ll}
\end{eqnarray}%
Multiplying Eq. (\ref{kk}) and Eq. (\ref{ll}), then we obtain the photon
arrival rate in optomecanical cavity, which is shown in Fig. 5. It can be
seen clearly that if the decay rate of the cavity which is considered in
\cite{Bouwmeester12} is small, i.e., $\omega _{m}=4\kappa $ (red line), the
photon arrival rate is almost distributed from $0$ to $2\pi $. Combined with
the amplifications from Fig. 3 within time $\omega _{m}t=2\pi $, the
implementation of our scheme is feasible in experiment. \newline

\section{Conclusion}

In conclusion, we use the quantum beam splitter in stead of the beam
splitter in the delayed-choice experiment. When combined with the weak
measurement and the quantum beam splitter play a role of the postselection,
it will be shown that the wave and particle behaviours of the measured
system give rise to a surprising amplification effect. We find that the
wave-particle duality of the measured system can be used as generation
mechanism for weak measurement amplification, that has not previously been
revealed. This amplification effect can't be explained by Maxwell's
equations \cite{Suter95}, due to the entanglement of the system and the
ancilla in quantum beam splitter. This amplification mechanism about the
wave-particle duality of the measured system lead us to a scheme for
implementation in optomechanical system. Moreover, the displacement of the
mirror's position in optomechanical system can appear within a large
evolution time zone, which is a counterintuitive negative amplification
effect. The results provide us an alternative method to discuss weak
measurement and also deepen our understanding of the weak measurement.

\section*{Acknowledgments}

This work was supported by the National Natural Science Foundation of China
under grants No. 11175033. Y. M. Y. acknowledges support from the National
Natural Science Foundation of China (Grant Nos. 61275215) and the National
Fundamental Research Program of China (Grant No. 2011CBA00203).

\end{document}